\documentclass[conference]{IEEEtran}
\IEEEoverridecommandlockouts
\usepackage{cite}
\usepackage{amsmath,amssymb,amsfonts}
\usepackage{algorithmic}
\usepackage{graphicx}
\usepackage{textcomp}
\usepackage{xcolor}
\usepackage{soul}

\usepackage[top=0.7in, left=0.680in , right=0.680in, bottom = 1.05in]{geometry}

\begin{document}
	
	\title{Secure Performance Analysis of RIS-aided Wireless Communication Systems}

	\author{\IEEEauthorblockN{Dinh-Thuan Do\textsuperscript{$*$}, 
			Anh-Tu Le\textsuperscript{$ \# $}, and Shahid Mumtaz\textsuperscript{$\dag $ }}
		
		\IEEEauthorblockA{\textsuperscript{$*$}\textit{Department of Electrical and Computer Engineering, The University of Texas at Austin, Austin, TX 78712, USA.}}

		\IEEEauthorblockA{\textsuperscript{$ \# $}\textit{Faculty of Information Technology,} \textit{Van Lang University,} Ho Chi Minh City, Vietnam}
		
		\IEEEauthorblockA{\textsuperscript{$\dag $} \textit{Instituto de Telecomunicações, Aveiro, Portugal}}
		\IEEEauthorblockA{\textsuperscript{$\dag $}{ARIES Research Center, Universidad Antonio de Nebrija, C/Pirineos, 55, E-28040, Madrid, Spain.}}
		
		\IEEEauthorblockA{Email: dodinhthuan@utexas.edu,  leanhtu@iuh.edu.vn and smumtaz@av.it.pt}
	}

	\maketitle
	
	\begin{abstract}
		Since Internet of Things (IoT)
		is suggested as the fundamental platform to adapt massive
		connections and secure transmission, we study physical-layer authentication in the point-to-point wireless systems relying on reconfigurable intelligent surfaces (RIS) technique. Due to lack of direct link from IoT devices (both
		legal and illegal devices) to the access point, we benefit from RIS by considering two main secure performance metrics. As main goal, we examine the secrecy
		performance of a RIS-aided
		wireless communication systems which show secure performance
		in the presence of an eavesdropping IoT devices. In this circumstance, RIS is placed between the access point and
		the legitimate devices and is designed to enhance the link security.
		To specify secure system performance metrics, we firstly present
		analytical results for the secrecy outage probability. Then, secrecy rate is further examined. Interestingly, we are to control both the average signal-to-noise ratio at the source and the number of metasurface
		elements of the RIS to achieve improved system performance. We verify derived expressions by matching Monte-Carlo simulation and analytical results.
	\end{abstract}
	
	
	\section{Introduction}
	
	By enabling interconnection with the Internet, advantages of Internet of Things (IoT) with various kinds of things (such as smart phones, Vehicles, sensors, etc.) need reliable and secure connections to provide promising applications including e-health, intelligent transportation systems, smart home and smart city \cite{b1}, \cite{b2}. In the perspective of live, work, and communication, IoT plays an important role to improve quality of experience. Unfortunately, IoT also causes an increasing challenge in term of its security with ubiquitous connected devices associated with our daily activities such as monitoring, recording and analyzing data \cite{b3}, \cite{b4}. In fact, since the malicious attackers occupy the opportunities to manipulate the physical objects in IoT system which cause severe problems both the cyber world and cyber-physical world \cite{b5}-\cite{b7}. For instance, several situations such as unlock the doors or disable the braking and steering Systems are related to vulnerabilities found in connected cars or hackers. Therefore, it is necessary to address urgently these security threats and issues to reduce crucial concerns for IoT technologies.
	
	There are lots of studies to tackle such threats. For example, the authors in \cite{b8} studied heterogeneous Internet of Things (IoT) and multi-access mobile edge computing by exploiting the secure wiretap coding, multi-node cooperation and resource allocation, signal processing. In addition, physical layer key generation and authentication are deployed to cope with the emerging security challenges. In particular, physical layer security (PLS) is now emerging as a promising candidate for providing perfect security, thus, there are many existing works related to PLS. To provide secure unmanned aerial vehicle (UAV) system can be used as jammers and adaptively adjusted the locations over time \cite{b9}. In cellular vehicle-to-everything (C-V2X) communication networks, investigate the potential of these PLS techniques is used to improve the security \cite{b10}. The authors proposed stochastic geometry approach to study the PLS of an artificial noise (AN) aided C-V2X network In work of \cite{b11}, device to device (D2D) communications and cellular networks are joint implement to show benefits of spectrum sharing with respect to reliability and robustness. To evaluate the cooperative system model, they derived formulas of closed-form expressions for the D2D outage probability, the secrecy outage probability, and the probability of non-zero secrecy capacity. The authors of \cite{b12} analyzed the secrecy performance in a wireless system by using a stochastic geometry approach. 
	
	Recently, to find more energy efficiency in transmission, reconfigurable intelligent surface (RIS)-aided transmission has been introduced with promising performance improvement. In principle, RIS consists a large number of low-cost reflecting elements, which is expected to improve the performance of wireless communication networks including security \cite{b13}, \cite{b14}. The programmable passive reflecting array is facilitated at RIS to reflect signals. To create better propagation conditions for the intended users, signals transmitted from the base station (BS) need the reflecting angle of the RIS intelligent adjusted to enhance extended coverage, transmission quality. As main advantages, RIS is known as a cost-effective transmission technique which does not require high implementation Costs and can be readily integrated into existing wireless communication system. Furthermore, low power and the limited spectrum are necessary to deploy RIS to provide more energy and spectrum efficient which is better in comparison with conventional relaying systems. In term of the spectral efficiency, RIS essentially works in the full-duplex mode without causing any interference and adding thermal noise . Moreover, by increasing the number of RIS elements, RIS-aided system can improve performance gains \cite{b13}. The single-user case has been well investigated in \cite{b17}, \cite{b18}. By joint precoding and IRS design, the RIS with multiuser case has also been investigated. By properly choosing the reflecting angles of the RIS elements, the existing works have shown that the RIS-aided systems are able to reduce transmission power at the BS \cite{b17}, \cite{b19}, improve the weighted sum-rate \cite{b20}, enhance performance gain, for instance the proposed NOMA RIS-aided system in \cite{b21}, \cite{b22} with 32 RIS elements has about 2.5 dB performance gain over the conventional massive multiple-input-multiple-output system with 64 transmit antennas \cite{b22}.

	To fill the gaps missing in above studies, it is necessary to investigate how hardware impairment \cite{Boulogeorgos2020} affects to secure performance of a point-to-point RIS-aided system. We design a simple transmission relying
	on PLS by enabling RIS in current IoT systems. Motivated by \cite{b13}, this paper aims to provide analytical results and to study secure performance of point-to-point RIS-aided wireless system with imperfect hardware.

	\section{System model}
	\begin{figure}[!h]
		\centering
		\includegraphics[width=0.5\textwidth]{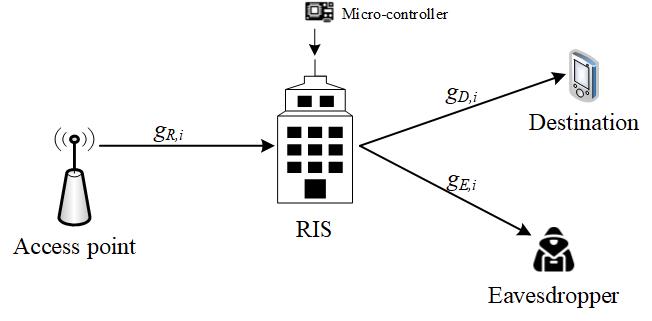}
		\caption{System model of point-to-point RIS considering PLS.}
		\label{model}
	\end{figure}
	
	We just focus on system performance of a downlink RIS. A pair of devices including base station or access point and destination is allowed to foster transmission efficiency with assistance of RIS which acts as a intermediate device. Such link is evaluated in term of secure performance since existence of an eavesdropper which tries to overhear signal from legal link. We need multiple metasurfaces elements to reflect signal to destination. These $ N $ metasurface elements are adjusted separately with different phases to optimal received signal at destination. In this scenario, an eavesdropper can not overhear signal from the access point which is likely provided by robust algorithm compared with RIS.

	In this section, we consider a access point (S), a destination, a eavesdropper (E) and a RIS with $N$ reflecting elements. Moreover, we assume direct-link transmission does not exist due to the blockage. Furthermore, we assume that the RIS can obtain the full channel state information (CSI) of the channels to realize the phase shift. To analyse system performance, we conduct computations related to information processing for distant device. We examine the received signal at intended user and eavesdropper respectively as
	\begin{equation} \label{Eq1}
y_D  = \sqrt {P_S } \sum\limits_{i = 1}^N {g_{R,i} g_{D,i} \xi _i } \left( {x + \eta _{D,t} } \right) + \eta _{D,r}  + n_D ,
	\end{equation}
	\begin{equation} \label{Eq2}
y_E  = \sqrt {P_S } \sum\limits_{i = 1}^N {g_{R,i} g_{E,i} \xi _i } \left( {x + \eta _{E,t} } \right) + \eta _{E,r}  + n_E ,
	\end{equation}
	where $x$ is the signal of S, $P_S$ stands for the transmit power at S, $\eta _{D,t}  \sim CN\left( {0,\kappa _{D,t}^2 } \right)$, $\eta _{D,r}  \sim CN\left( {0,X_1^2 \kappa _{D,r}^2 P_S } \right)$, $\eta _{E,t}  \sim CN\left( {0,\kappa _{E,t}^2 } \right)$, $\eta _{E,r}  \sim CN\left( {0,X_2^2 \kappa _{E,r}^2 P_S } \right)$ are the denotes the distortion noise from residual hardware impairment as \cite{Boulogeorgos2020}, $\kappa _{D,t}^2$, $\kappa _{D,r}^2$, $\kappa _{E,t}^2$, $\kappa _{E,r}^2 $ are the levels of hardware impairment, $n_D$ and $n_E$ are assumed as additive white Gaussian noise (AWGN) with ${\cal{CN}}(0,N_0)$. For characterizations of RIS, $\xi _i  = \varpi _i \left( {\phi _i } \right)e^{j\phi _i }$ is called as the reflecting coefficient corresponding to the $i$-th reflector of the RIS. We treat perfect RIS elements with $\varpi _i \left( {\phi _i } \right)$ is considered as the ideal phase shifts $(i = 1,...,N)$. It is noted that channel gains denoted by ${g_{R,i}}$, ${g_{D,i}}$, ${g_{E,i}}$ and they follow $g_{R,i}  = {\sqrt {d_{SR} } } ^{ - \chi } f_{R,i} e^{ - j\theta _i } $, $g_{D,i}  = {\sqrt {d_{RD} } } ^{ - \chi } f_{D,i} e^{ - j\varphi _i } $ and $g_{E,i}  = {\sqrt {d_{RE} } } ^{ - \chi } f_{E,i} e^{ - j\delta _i } $, in which distances $d_{SR}$, $d_{RD}$ and $d_{RE}$ correspond to link between $S$-RIS, RIS-$D$ and RIS-$E$, respectively, $\theta_i$, $\varphi_i$ and $\delta_i$ represent for the phases of the channel gains while $\chi$ is denoted as the path loss coefficient, the amplitudes of channels are$f_{R,i} ,f_{D,i}, f_{E,i}$ which follow independent distributed Rayleigh random variables (RVs). The full knowledge of the phases of channel gain are available at receivers. We then calculate the signal to noise and distortion rate (SNDR) at $D$ as below
	\begin{equation} \label{Eq3}
\gamma _D  = \frac{{\frac{{P_S }}
{{\left( {d_{SR} d_{RD} } \right)^\chi  }}\left| {\sum\limits_{i = 1}^N {f_{R,i} f_{D,i} e^{j\varphi _i  - j\phi _i  - j\theta _i } } } \right|^2 }}
{{\frac{{\left( {\kappa _{D,t}^2  + \kappa _{D,r}^2 } \right)P_S }}
{{\left( {d_{SR} d_{RD} } \right)^\chi  }}\left| {\sum\limits_{i = 1}^N {f_{R,i} f_{D,i} e^{j\varphi _i  - j\phi _i  - j\theta _i } } } \right|^2  + N_0 }}.
	\end{equation}
	
	As upper bound of such SNDR, \eqref{Eq2} can be replaced by $\phi_i = \theta_i + \varphi_i$ \cite{Basar2019}. Then, we want to maximize of $\gamma_D$ as
	\begin{equation} \label{Eq4}
\begin{gathered}
  \gamma _D  = \frac{{\frac{{P_S }}
{{\left( {d_{SR} d_{RD} } \right)^\chi  }}\left| {\sum\limits_{i = 1}^N {f_{R,i} f_{D,i} } } \right|^2 }}
{{\frac{\left( {\kappa _{D,t}^2  + \kappa _{D,r}^2 } \right){P_S }}
{{\left( {d_{SR} d_{RD} } \right)^\chi  }}\left| {\sum\limits_{i = 1}^N {f_{R,i} f_{D,i} } } \right|^2  + N_0 }} \hfill \\
   = \frac{{\rho _D }}
{{\left( {\kappa _{D,t}^2  + \kappa _{D,r}^2 } \right)\rho _D  + 1}}, \hfill \\ 
\end{gathered} 
	\end{equation}
	where $X_1  = \left| \sum\limits_{i = 1}^N {f_{R,i} f_{D,i} } \right|$, $
	\bar \gamma _{D}  = \frac{{P_S }}{{\left( {d_{SR} d_{RD} } \right)^\chi  N_0 }}$ denotes the average SNR and $\rho _D  = \bar \gamma _D X_1^2 $. 
	
	In the eavesdropper link, due to the phase shifts $\phi_i$ are designed to compensate for the effect of the fading channel coefficients of the main link so the residual phase of the eavesdropper link will be much larger than the legitimate counterpart and whenever $\delta _{i } \sim  \mathcal{U}[-\pi:\pi)$ regardless of the phase distribution of $\phi _i$ \cite{Vega2021}. Thus, the SNR at E is given by
	\begin{equation} \label{Eq5}
\begin{gathered}
  \gamma _E  = \frac{{\frac{{P_S }}
{{\left( {d_{SR} d_{RE} } \right)^\chi  }}\left| {\sum\limits_{i = 1}^N {f_{R,i} f_{E,i} } } \right|^2 }}
{{\frac{{\left( {\kappa _{E,t}^2  + \kappa _{E,r}^2 } \right)P_S }}
{{\left( {d_{SR} d_{RE} } \right)^\chi  }}\left| {\sum\limits_{i = 1}^N {f_{R,i} f_{E,i} } } \right|^2  + N_0 }} \hfill \\
   = \frac{{\rho _E }}
{{\left( {\kappa _{E,t}^2  + \kappa _{E,r}^2 } \right)\rho _E  + 1}}, \hfill \\ 
\end{gathered} 
	\end{equation}
	where $X_2  = \left|\sum\limits_{i = 1}^N {f_{R,i} f_{E,i} }\right| $, $\bar \gamma _E  = \frac{{P_S }}{{\left( {d_{SR} d_{RE} } \right)^\chi  N_0 }}$ is the average SNR and $\rho _E  = \bar \gamma _E X_2^2 $. The maximum achievable secrecy capacity is defined by \cite{Bloch2008}
	\begin{equation} \label{Eq6}
		R_S  = \max \left\{ {\log _2 \left( {1 + \gamma _D } \right) - \log _2 \left( {1 + \gamma _E } \right),0} \right\}. 
	\end{equation}
	
	\section{Secrecy Performance Analysis}
	The point-to-point RIS-aided IoT system is characterized secure performance by evaluating secure outage probability and secrecy rate. 	
	\subsection{Channel model}
	We refer to channel model for $\gamma_D$. Starting from the probability density function (PDF) of $X^2_1$ \cite{Yang2020}
	\begin{equation} \label{Eq7}
		f_{X_1^2 } \left( x \right) = \frac{1}
		{{2\sigma ^2 }}\left( {\frac{x}
			{\lambda }} \right)^{ - \frac{1}
			{4}} e^ { { - \frac{{x + \lambda }}
				{{2\sigma ^2 }}} }I_{ - \frac{1}
			{2}} \left( {\frac{{\sqrt {x\lambda } }}
			{{\sigma ^2 }}} \right),
	\end{equation}
	where $\lambda = (\frac{N\pi}{4})^2$, $\sigma^2 = N(1 - \frac{\pi^2}{16})^2$ and $I_{\nu}(\bullet )$ represents for the Bessel functions of the first class with order $\nu$ \cite{Book}. Then we have the PDF of $\rho_D$ as
	\begin{equation} \label{Eq8}
		f_{\rho_D } \left( x \right) = \frac{1}
		{{2\sigma ^2 }}\left( {\frac{x}
			{\lambda }} \right)^{ - \frac{1}
			{4}} \left( {\frac{1}
			{{\bar \gamma _D }}} \right)^{\frac{3}
			{4}} e^{ - \frac{{x + \lambda \bar \gamma_r }}
			{{2\bar \gamma _r \sigma ^2 }}} I_{ - \frac{1}
			{2}} \left( {\frac{{\sqrt {x\lambda } }}
			{{\sqrt {\bar \gamma _D } \sigma ^2 }}} \right).
	\end{equation}
	
	Based on \cite[Eq. 8.445]{Book}, we have
	\begin{equation} \label{Eq9}
		I_v \left( z \right) = \sum\limits_{k = 0}^\infty  {} \frac{1}
		{{k!\Gamma \left( {v + k + 1} \right)}}\left( {\frac{z}
			{2}} \right)^{v + 2k}. 
	\end{equation}
	
	The, we can rewrite \eqref{Eq8} as
	\begin{equation} \label{Eq10}
		f_{\rho_D} \left( x \right) = \sum\limits_{k = 0}^\infty  {\frac{{\lambda ^k e^{ - \frac{{\lambda }}
						{{2\sigma ^2 }}} x^{k - \frac{1}
						{2}} e^{ - \frac{x}
						{{2\bar \gamma _D \sigma ^2 }}} }}
			{{k!\Gamma \left( {k + \frac{1}
						{2}} \right)\left( {2\sigma ^2 } \right)^{2k+\frac{1}
						{2}} \left( {\bar \gamma _D } \right)^{k + \frac{1}
						{2}} }}}. 
	\end{equation}
	
	Let denote $Q_m(a;b)$ ass the Marcum Q-function \cite{Kapinas2009} and $\gamma(\bullet,\bullet)$ ass the lower gamma incomplete \cite{Book}, the cumulative distribution function (CDF) of $\gamma_D$ is expressed by
	\begin{equation} \label{Eq11}
		\begin{gathered}
			F_{\rho_D} \left( x \right) = 1 - Q_{\frac{1}
				{2}} \left( {\frac{{\sqrt \lambda  }}
				{\sigma },\sqrt {\frac{x }
					{{\bar \gamma _D \sigma ^2 }}} } \right) \hfill \\
			= \sum\limits_{k = 0}^\infty  {\frac{{e^{ - \frac{\lambda }
							{{2\sigma ^2 }}} }}
				{{k!\Gamma \left( {k + \frac{1}
							{2}} \right)}}} \left( {\frac{\lambda }
				{{2\sigma ^2 }}} \right)^k \gamma \left( {k + \frac{1}
				{2},\frac{x}
				{{2\bar \gamma _D \sigma ^2 }}} \right). \hfill \\ 
		\end{gathered} 
	\end{equation}

	We treat the channel model of $\gamma_E$ with the phase shifts of illegal link adjusted by the legitimate link, Rayleigh distribution is adopted for $X_2$ \cite{Vega2021} with variance $E\{X^2_2\} = 1/n$ to give the PDF of $\rho _E $ as
	\begin{equation} \label{Eq12}
		f_{\rho _E }  = \frac{1}
		{{\lambda _E }}e^{ - \frac{x}
			{{\lambda _E }}} ,
	\end{equation}
	where $\lambda _E  = \bar \gamma _E N$.
	
	\subsection{Secure Outage Probability}
	By denoting $C_{th}$ as the target secrecy rate, the secure outage probability (SOP) is given by
	\begin{equation} \label{Eq13}
		SOP = \Pr \left( {R_S  < C_{th} } \right) = \Pr \left( {\frac{{1 + \gamma _D }}
			{{1 + \gamma _E }} < \gamma _{th} } \right),
	\end{equation}
	where $\gamma _{th}  = 2^{C_{th} }$, . 
	
	\textbf{Proposition 1:} We compute SOP in the closed-form as \eqref{Eq15}, shown at the top of the next page.
	\begin{table*}
		\begin{equation} \label{Eq15}
SOP \approx \sum\limits_{k = 0}^\infty  {\frac{{e^{ - \frac{\lambda }
{{2\sigma ^2 }}} \theta _3 \pi }}
{{\lambda _E k!\Gamma \left( {k + \frac{1}
{2}} \right)2\theta _2 N}}} \left( {\frac{\lambda }
{{2\sigma ^2 }}} \right)^k \sum\limits_{n = 1}^N {\sqrt {1 - \varphi _n^2 } } \gamma \left( {k + \frac{1}
{2},\frac{{\theta _1 \theta _3 \left( {1 + \varphi _n } \right) + \theta _2 \vartheta }}
{{2\bar \gamma _D \sigma ^2 \theta _2 \left( {\theta _3  - \theta _3 \left( {1 + \varphi _n } \right)} \right)}}} \right)e^{ - \frac{{\theta _3 \left( {1 + \varphi _n } \right)}}
{{\theta _2 \lambda _E }}}.
		\end{equation}
		\hrulefill
	\end{table*}
	
	\textbf{Proof:} See Appendix A.
	
	To look insights of secure performance at high SNR regime, it can be approximated the SOP of \eqref{Eq13} as
	\begin{equation} \label{Eq16}
\begin{gathered}
  SOP^\infty   \approx \Pr \left( {\frac{{\gamma _D }}
{{\gamma _E }} < \gamma _{th} } \right) \hfill \\
   \approx \Pr \left( {\rho _D  < \frac{{\gamma _{th} \rho _E }}
{{1 - \theta _4 \rho _E }}} \right) \hfill \\
   \approx \int\limits_0^\infty  {F_{\gamma _D } \left( {\frac{{\gamma _{th} x}}
{{1 - \theta _4 x}}} \right)f_{\gamma _E } \left( x \right)} dx. \hfill \\ 
\end{gathered} 
	\end{equation}
where $\theta _4  = \gamma _{th} \left( {\kappa _{D,t}^2  + \kappa _{D,r}^2 } \right) - \left( {\kappa _{E,t}^2  + \kappa _{E,r}^2 } \right)$. Then, we can write $SOP^{\infty}$ with the help of \eqref{Eq11} and \eqref{Eq12} as
	\begin{equation} \label{Eq17}
\begin{gathered}
  SOP^\infty   \approx \sum\limits_{k = 0}^\infty  {\frac{{e^{ - \frac{\lambda }
{{2\sigma ^2 }}} }}
{{\lambda _E k!\Gamma \left( {k + \frac{1}
{2}} \right)}}} \left( {\frac{\lambda }
{{2\sigma ^2 }}} \right)^k  \hfill \\
   \times \int\limits_0^{\frac{1}
{{\theta _4 }}} {\gamma \left( {k + \frac{1}
{2},\frac{{\gamma _{th} x}}
{{2\bar \gamma _D \sigma ^2 \left( {1 - \theta _4 s} \right)}}} \right)e^{ - \frac{x}
{{\lambda _E }}} } dx. \hfill \\ 
\end{gathered} 
	\end{equation}
	
	Eventually, $SOP^{\infty}$ is expressed by
	\begin{equation} \label{Eq18}
\begin{gathered}
  SOP^\infty   \approx \sum\limits_{k = 0}^\infty  {\frac{{e^{ - \frac{\lambda }
{{2\sigma ^2 }}} }}
{{\lambda _E k!\Gamma \left( {k + \frac{1}
{2}} \right)}}} \left( {\frac{\lambda }
{{2\sigma ^2 }}} \right)^k \frac{\pi }
{{2\theta _4 N}}\sum\limits_{n = 0}^N {\sqrt {1 - \varphi _n } }  \hfill \\
   \times \gamma \left( {k + \frac{1}
{2},\frac{{\gamma _{th} \left( {1 + \varphi _n } \right)}}
{{2\theta _4 \bar \gamma _D \sigma ^2 \left( {2 - \left( {1 + \varphi _n } \right)} \right)}}} \right)e^{ - \frac{{\left( {1 + \varphi _n } \right)}}
{{\theta _4 \lambda _E }}}.  \hfill \\ 
\end{gathered} 
	\end{equation}

	\subsection{Analysis of Secrecy capacity}
	Next, we characterize the average secrecy capacity as \cite{Badarneh2018}
	\begin{equation} \label{Eq19}
		\begin{gathered}
			\bar R_S  = \mathbb{E}\left[ {R_S } \right] \hfill \\
			= \mathbb{E}\left[ {\log _2 \left( {1 + \gamma _D } \right)} \right] 
			- \mathbb{E}\left[ {\log _2 \left( {1 + \gamma _E } \right)} \right]. \hfill \\ 
		\end{gathered} 
	\end{equation}
	
	\textbf{Proposition 2:} The closed-form of $\bar R_S$ is given as \eqref{Eq20}, shown at the top of the next page.
	\begin{table*}
		\begin{equation} \label{Eq20}
\begin{gathered}
  \bar R_S  \approx \frac{1}
{{\ln \left( 2 \right)}}\sum\limits_{k = 0}^\infty  {\frac{{e^{ - \frac{\lambda }
{{2\sigma ^2 }}} }}
{{k!\Gamma \left( {k + \frac{1}
{2}} \right)}}} \left( {\frac{\lambda }
{{2\sigma ^2 }}} \right)^k \frac{\pi }
{N}\sum\limits_{n = 1}^N {\sqrt {1 - \varphi _n^2 } } \frac{{\Gamma \left( {k + \frac{1}
{2},\frac{{\left( {1 + \varphi _n } \right)}}
{{2\bar \gamma _D \sigma ^2 \left( {2 - \left( {1 + \varphi _n } \right)} \right)\left( {\kappa _{D,t}^2  + \kappa _{D,r}^2 } \right)}}} \right)}}
{{2\left( {\kappa _{D,t}^2  + \kappa _{D,r}^2 } \right) + \left( {1 + \varphi _n } \right)}} \hfill \\
   + \frac{{e^{\frac{1}
{{\lambda _E \left( {\kappa _{E,t}^2  + \kappa _{E,r}^2 } \right)}}} }}
{{\ln \left( 2 \right)}}\left[ {Ei\left( { - \frac{1}
{{\left( {\kappa _{E,t}^2  + \kappa _{E,r}^2 } \right)\lambda _E }} - \frac{1}
{{\lambda _E }}} \right) - Ei\left( { - \frac{1}
{{\lambda _E }}} \right)} \right]. \hfill \\ 
\end{gathered} 
		\end{equation}
		\hrulefill
	\end{table*}
	
	\textbf{Proof:} See Appendix B

	\section{Simulation result}
	To confirm exactness of mathematical expressions derived in the paper, we provide lots of numerical results. Except for specific case, we set $C_{th} = 1$. For hardware impairment, we set $\kappa^2 = \kappa_{D,t}=\kappa_{D,r}=\kappa_{E,t}=\kappa_{E,r}=0.01$. 
	
	\begin{figure}[!h]
		\centering
		\includegraphics[width=0.5\textwidth]{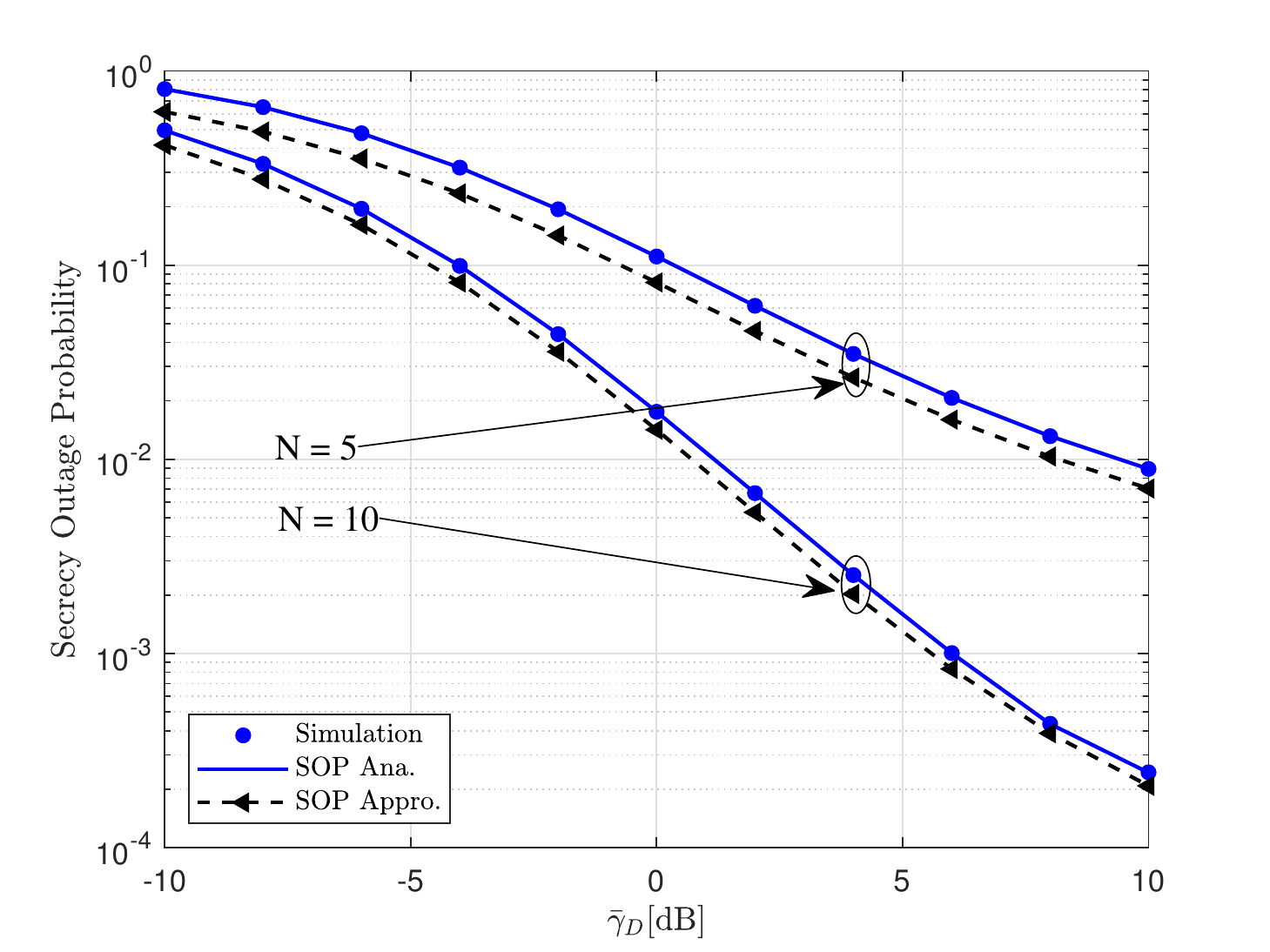}
		\caption{The SOP versus $\bar \rho_D$ varying $N$.}
		\label{fig1}
	\end{figure}
	
	In Fig. 2, we compare SOP performance for two cases of the number of RIS elements, i.e. $ N=5, 10 $. It can be confirmed that the transmit SNR at the access point $\bar\gamma_{D}$ is main factor affecting to SOP performance.In particular, high value of $\bar\gamma_{D}= 10$ leads to significant improvement of SOP. Our computation is checked by matching Monte-Carlo and analytical simulations, shown as consistence of related curves of SOP. Of course, higher value of $ N $ results in better SOP performance. For evaluation of approximate curves, it should be more precisely consider the case of $ N=10 $, i.e. two ways of computations for SOP are similar result. It can be explained that higher RIS elements makes significant improvement to related SNDR and SOP performance can be improved. Similarly, SOP performance depends on how eavesdropper makes influence to main link. Obviously, $\bar\gamma_{E}= -10$ is the best case among three cases of $\bar\gamma_{E}$. 
	
	\begin{figure}[!h]
		\centering
		\includegraphics[width=0.5\textwidth]{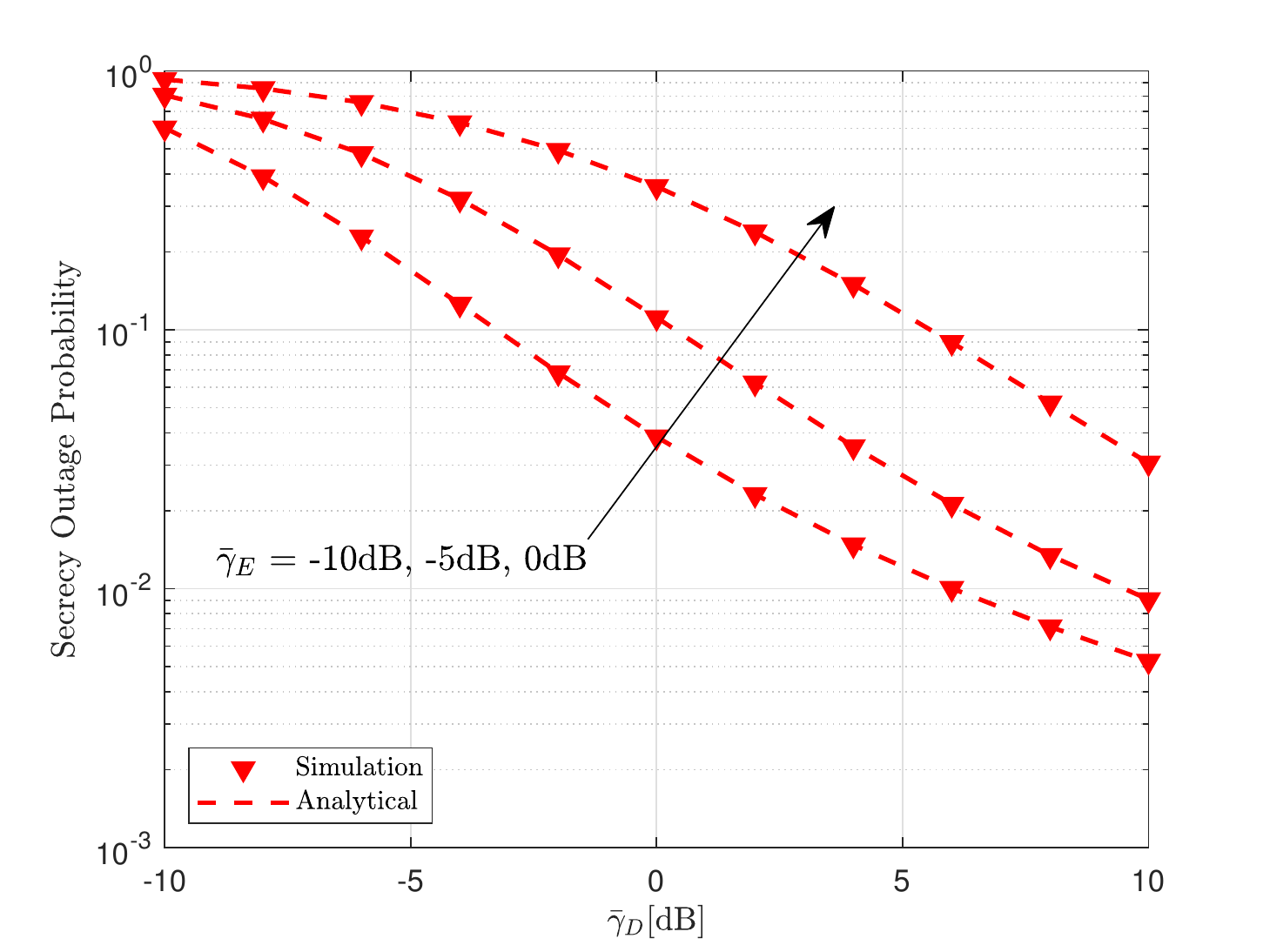}
		\caption{The SOP versus $\bar \rho_D$ varying $\bar \rho_E$ with $N=5$.}
		\label{fig2}
	\end{figure}

	\begin{figure}[!h]
		\centering
		\includegraphics[width=0.5\textwidth]{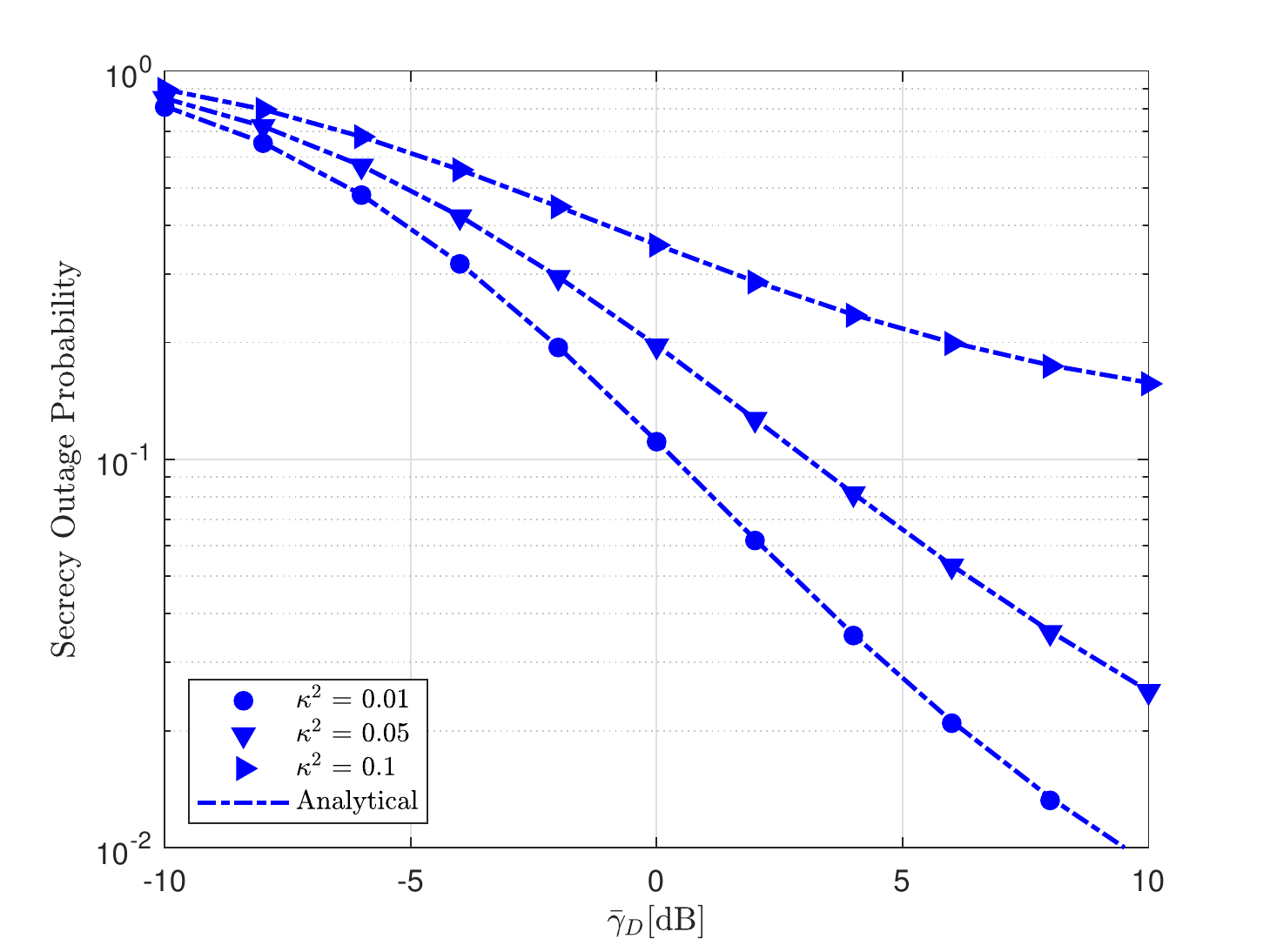}
		\caption{The SOP versus $\bar \rho_D$ varying $\kappa^2$ with $N=5$ and $\bar \rho_E$ = -10dB.}
		\label{fig3}
	\end{figure}
	
	\begin{figure}[!h]
		\centering
		\includegraphics[width=0.5\textwidth]{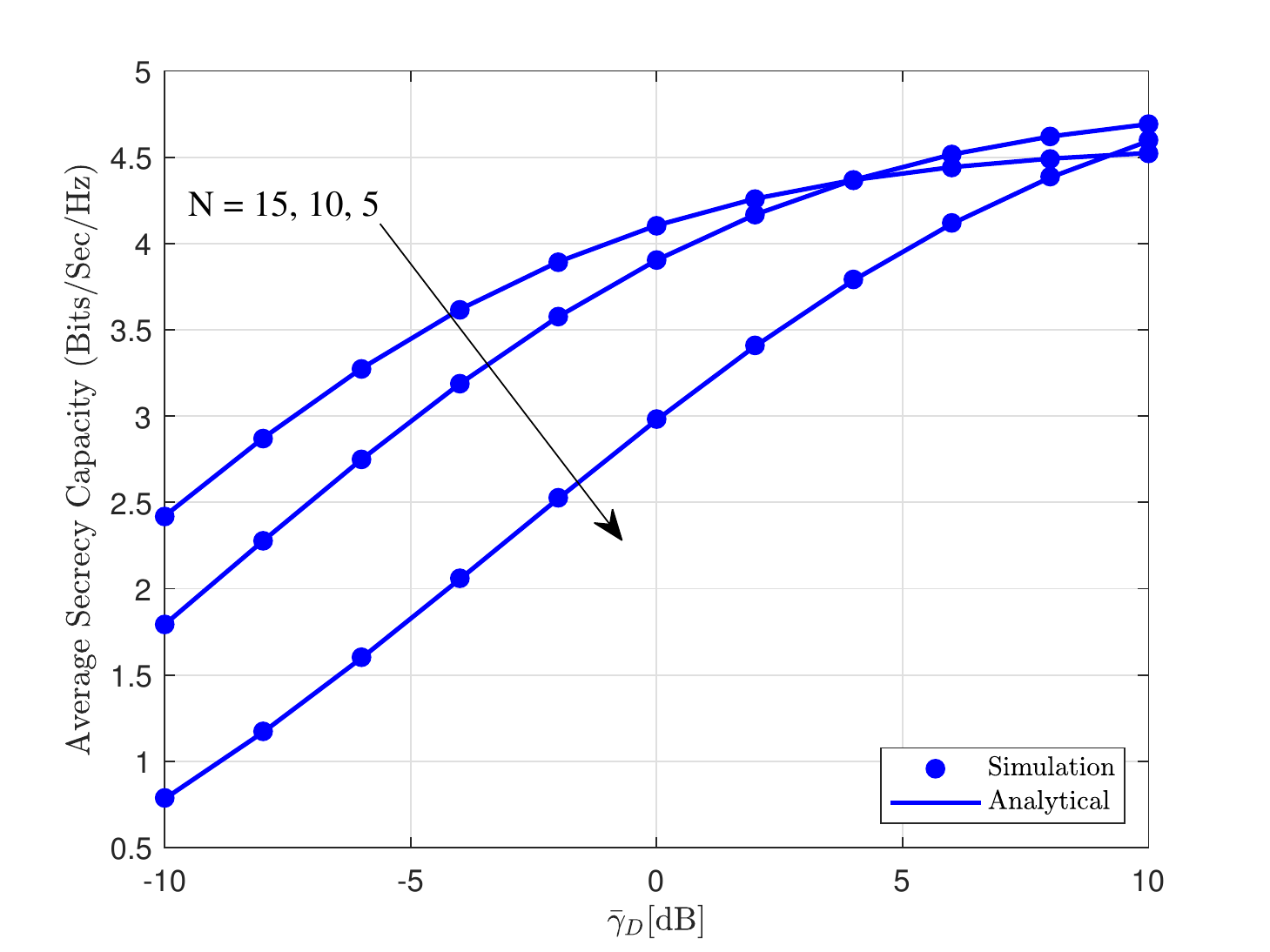}
		\caption{The average secrecy capacity versus $\bar \rho_D$ varying N with $\bar \rho_E$ = -10dB.}
		\label{fig4}
	\end{figure}

	\begin{figure}[!h]
		\centering
		\includegraphics[width=0.5\textwidth]{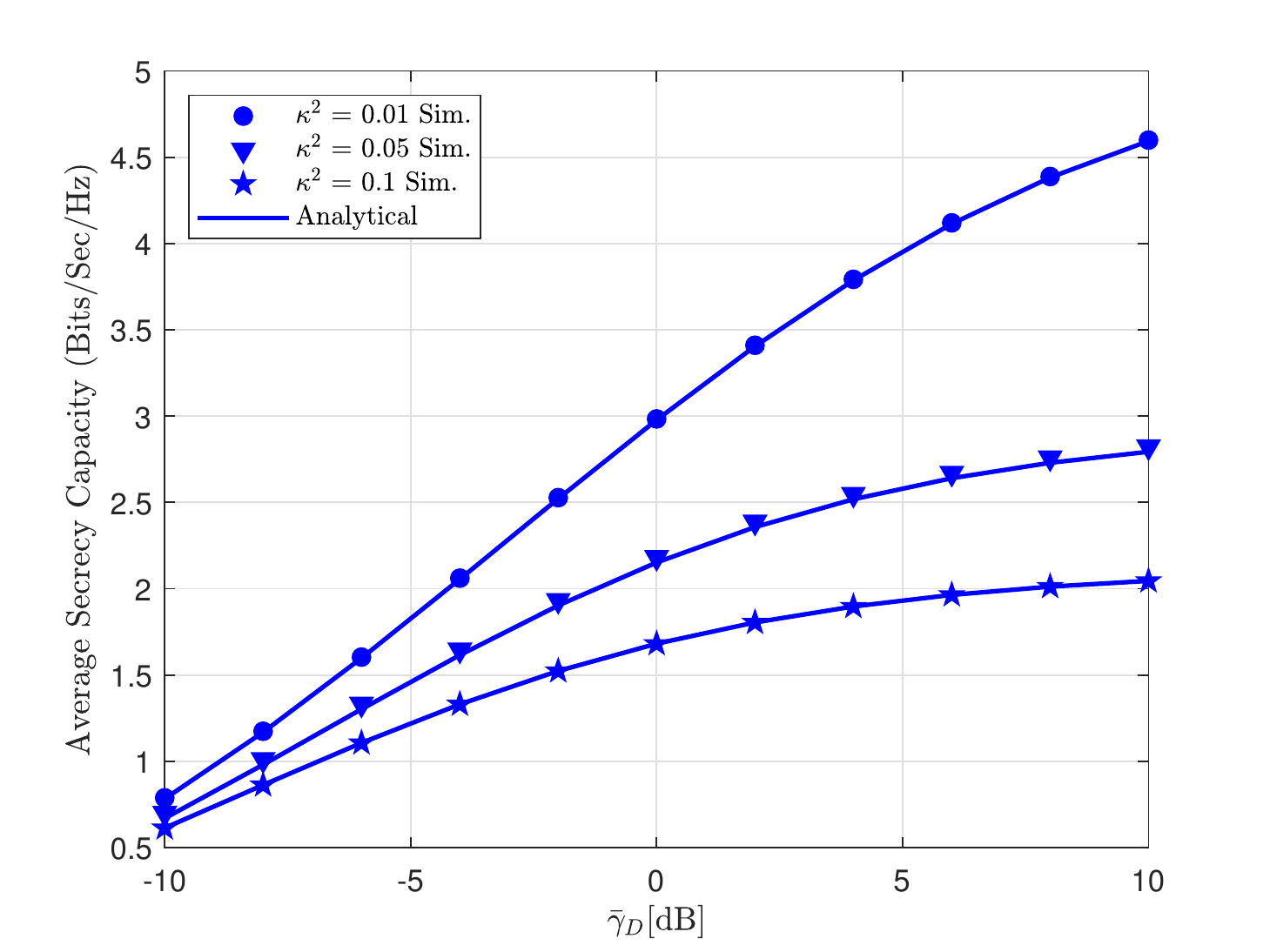}
		\caption{The average secrecy capacity versus $\bar \rho_D$ varying $\kappa^2$ with $N=5$ and $\bar \rho_E $ = -10dB.}
		\label{fig4}
	\end{figure}

	In Fig. 4, levels of hardware impairments lead to high impact on SOP performance. Especially, $\kappa^2=0.1$ is reported as worst case. Once imperfect hardware raises in system, we need to consider advanced technique as RIS to compensate such degraded factor.
	
	For secrecy capacity, we evaluate three cases of metasurfaces $ N $ which exhibit clear gaps among curves, shown as Fig. 5. In such way, by improving SNR at the access point $\bar \gamma_{D}$, we expect secrecy capacity would be better. It is worth noting that gaps among three curves are not clear at high value of $\bar\gamma_{D}$. In the view of levels of imperfect hardware, we see in Fig. 6 that lower value of $\kappa^2$ results in remarkable improvement of average secrecy capacity at the point $\bar \gamma_{D}= 10 (dB)$. 
	
%

	\section{Conclusion}
	The secure performance at devices has been evaluated in two main system metrics, i.e. secure outage probability and average secrecy capacity. We found that hardware impairments raise crucial concern on our study of secure performance. It should be put our awareness on structure of RIS when higher number of metasurfaces leads to better SNDR and corresponding secure performance can be improved. Although our study provides initial scenario of an point-to-point IoT deployment under advanced design of RIS, our analysis can be extended to more complicated scenarios such as multiple destinations, multiple eavesdroppers and multiple RISs.

	\section*{Appendix A}
	Substituting \eqref{Eq4} and \eqref{Eq5} into \eqref{Eq13}, we can rewrite SOP as \eqref{EqA1}, shown at the top of the next page, in which $\vartheta  = \gamma _{th}  - 1$.
	
	\begin{table*}
	\begin{equation} \label{EqA1}
SOP = \Pr \left( {\frac{{\rho _D }}
{{\left( {\kappa _{D,t}^2  + \kappa _{D,r}^2 } \right)\rho _D  + 1}} < \vartheta  + \frac{{\gamma _{th} \rho _E }}
{{\left( {\kappa _{E,t}^2  + \kappa _{E,r}^2 } \right)\rho _E  + 1}}} \right).
	\end{equation}
	\hrulefill
	\end{table*}
	
Then, we can rewrite \eqref{EqA1} as
	\begin{equation} \label{EqA2}
\begin{gathered}
  SOP = \Pr \left( {\rho _D  < \frac{{\theta _1 \rho _E  + \vartheta }}
{{\theta _3  - \theta _2 \rho _E }}} \right) \hfill \\
   = \int\limits_0^{\frac{{\theta _3 }}
{{\theta _2 }}} {f_{\rho _E } \left( x \right)F_{\rho _D } \left( {\frac{{\theta _1 x + \vartheta }}
{{\theta _3  - \theta _2 x}}} \right)dx.}  \hfill \\ 
\end{gathered} 
	\end{equation}

$\theta _1  = \vartheta \left( {\kappa _{E,t}^2  + \kappa _{E,r}^2 } \right) + \gamma _{th}$, \\
$\theta _2  = \vartheta \left( {\kappa _{E,t}^2  + \kappa _{E,r}^2 } \right)\left( {\kappa _{D,t}^2  + \kappa _{D,r}^2 } \right) + \gamma _{th} \left( {\kappa _{D,t}^2  + \kappa _{D,r}^2 } \right) - \left( {\kappa _{E,t}^2  + \kappa _{E,r}^2 } \right)$, \\
$\theta _3  = 1 - \vartheta \left( {\kappa _{D,t}^2  + \kappa _{D,r}^2 } \right)$.

Putting \eqref{Eq11} and \eqref{Eq12} into \eqref{EqA2}, we have
	\begin{equation} \label{EqA3}
\begin{gathered}
  SOP = \sum\limits_{k = 0}^\infty  {\frac{{e^{ - \frac{\lambda }
{{2\sigma ^2 }}} }}
{{\lambda _E k!\Gamma \left( {k + \frac{1}
{2}} \right)}}} \left( {\frac{\lambda }
{{2\sigma ^2 }}} \right)^k  \hfill \\
   \times \int\limits_0^{\frac{{\theta _3 }}
{{\theta _2 }}} {\gamma \left( {k + \frac{1}
{2},\frac{{\theta _1 x + \vartheta }}
{{\left( {\theta _3  - \theta _2 x} \right)2\bar \gamma _D \sigma ^2 }}} \right)e^{ - \frac{x}
{{\lambda _E }}} dx} . \hfill \\ 
\end{gathered} 
	\end{equation}

Then using Gaussian Chebyshev with $\varphi _n  = \cos \left( {\frac{{2n - 1}}{{2N}}\pi } \right)$. SOP can be obtained as \eqref{Eq15}.
		
	The proof is completed.
	
	\section*{Appendix B}
	
	First, we denote the first and second term of \eqref{Eq19} are $R_D$ and $R_E$ respectively. Then, the term $R_r$ with $r\in\{D,E\}$ can be calculated as
	\begin{equation} \label{EqB1}
		R_r  = \frac{1}
		{{\ln \left( 2 \right)}}\int\limits_0^\infty  {\frac{{1 - F_{\gamma _r } \left( x \right)}}
			{{1 + x}}} dx.
	\end{equation}
	
Then, the CDF of $\gamma_D$ can be written by
		\begin{equation} \label{EqB1A}
\begin{gathered}
  F_{\gamma _D } \left( x \right) = 1 - \sum\limits_{k = 0}^\infty  {\frac{{e^{ - \frac{\lambda }
{{2\sigma ^2 }}} }}
{{k!\Gamma \left( {k + \frac{1}
{2}} \right)}}} \left( {\frac{\lambda }
{{2\sigma ^2 }}} \right)^k  \hfill \\
   \times \Gamma \left( {k + \frac{1}
{2},\frac{x}
{{2\bar \gamma _D \sigma ^2 \left( {1 - x\left( {\kappa _{D,t}^2  + \kappa _{D,r}^2 } \right)} \right)}}} \right). \hfill \\ 
\end{gathered} 
	\end{equation}
	
	Next, putting \eqref{EqB1A} into \eqref{EqB1} we get $R_D$ as
	\begin{equation} \label{EqB2}
\begin{gathered}
  R_D  = \frac{1}
{{\ln \left( 2 \right)}}\sum\limits_{k = 0}^\infty  {\frac{{e^{ - \frac{\lambda }
{{2\sigma ^2 }}} }}
{{k!\Gamma \left( {k + \frac{1}
{2}} \right)}}} \left( {\frac{\lambda }
{{2\sigma ^2 }}} \right)^k  \hfill \\
   \times \int\limits_0^{\frac{1}
{{\left( {\kappa _{D,t}^2  + \kappa _{D,r}^2 } \right)}}} {\frac{{\Gamma \left( {k + \frac{1}
{2},\frac{x}
{{2\bar \gamma _D \sigma ^2 \left( {1 - x\left( {\kappa _{D,t}^2  + \kappa _{D,r}^2 } \right)} \right)}}} \right)}}
{{1 + x}}} dx. \hfill \\ 
\end{gathered} 
	\end{equation}
	
Similarly, using Gaussian Chebyshev, it can be obtained $R_D$ as

	%
	%
	%
	
	\begin{equation} \label{EqB5}
\begin{gathered}
  R_D  \approx \frac{1}
{{\ln \left( 2 \right)}}\sum\limits_{k = 0}^\infty  {\frac{{e^{ - \frac{\lambda }
{{2\sigma ^2 }}} }}
{{k!\Gamma \left( {k + \frac{1}
{2}} \right)}}} \left( {\frac{\lambda }
{{2\sigma ^2 }}} \right)^k \frac{\pi }
{N}\sum\limits_{n = 1}^N {\sqrt {1 - \varphi _n^2 } }  \hfill \\
   \times \frac{{\Gamma \left( {k + \frac{1}
{2},\frac{{\left( {1 + \varphi _n } \right)}}
{{2\bar \gamma _D \sigma ^2 \left( {2 - \left( {1 + \varphi _n } \right)} \right)\left( {\kappa _{D,t}^2  + \kappa _{D,r}^2 } \right)}}} \right)}}
{{2\left( {\kappa _{D,t}^2  + \kappa _{D,r}^2 } \right) + \left( {1 + \varphi _n } \right)}}. \hfill \\ 
\end{gathered} 
	\end{equation}
	
	Then, the term $R_E$ is calculated as
	\begin{equation} \label{EqB6}
R_E  = \frac{1}
{{\ln \left( 2 \right)}}\int\limits_0^{\frac{1}
{{\left( {\kappa _{E,t}^2  + \kappa _{E,r}^2 } \right)}}} {\frac{{e^{ - \frac{x}
{{\lambda _E }}} }}
{{1 + x}}} dx.
	\end{equation}
	
	By relying on result in \cite[Eq. 3.352.1]{Book}, the closed-form of $R_E$ can be obtained as
	\begin{equation} \label{EqB7}
\begin{gathered}
  R_E  = \frac{{e^{\frac{1}
{{\lambda _E }}} }}
{{\ln \left( 2 \right)}}\left[ { - Ei\left( { - \frac{1}
{{\lambda _E }}} \right)} \right. \hfill \\
   + \left. {Ei\left( { - \frac{1}
{{\lambda _E \left( {\kappa _{E,t}^2  + \kappa _{E,r}^2 } \right)}} - \frac{1}
{{\lambda _E }}} \right)} \right] .\hfill \\ 
\end{gathered} 
	\end{equation}
	
	Putting \eqref{EqB5} and \eqref{EqB7} into \eqref{Eq19} we obtain \eqref{Eq20}. The proof is completed.

	\vspace{12pt}
	
\end{document}